# Vortex gyration mediated by spin waves driven by an out-of-plane oscillating magnetic field

Yan Liu[*], Huanan Li, Yong Hu, and An Du

*College of Sciences, Northeastern University, Shenyang 110819, China*

Corresponding author: Yan Liu
E-mail: liuyanphys@mail.neu.edu.cn

**Abstract:**

In this letter we address the vortex core dynamics involved in gyration excitation and damping change by out-of-plane oscillating magnetic fields. When the vortex core is at rest under the effect of in-plane bias magnetic fields, the spin waves excited by the perpendicular magnetic field can induce obvious vortex gyration. When simultaneously excite spin waves and vortex gyrotropic motion, the gyration damping changes. Analysis of the system energy allows us to explain the origin of the spin-wave-mediated vortex gyration.

## Introduction

The magnetic ground state in micro- and nanosized circular magnetic disks is the vortex state, which is characterized by an in-plane curling magnetization, and a perpendicularly magnetized core at the center of the disk [1]. A vortex state is commonly identified by two parameters: polarity of vortex core, pointing up ($p=1$) or down ($p=-1$), and chirality of the in-plane magnetization, clockwise ($c=-1$) or counterclockwise ($c=1$). Currently, precise knowledge of the vortex dynamics is important due to their applications in many proposed spintronics devices, from extremely sensitive magnetic field sensors, to spin polarized current-tunable microwave vortex nano-oscillators and vortex magnetic random access memory [2-4].

The vortex in nanodisks possesses two types of dynamic spin excitation modes, the low frequency gyrotropic mode and the higher frequency spin-wave mode. The gyrotropic mode is characterized by spiral-like core motion around its equilibrium position. Its frequency generally is in the range of hundreds of megahertz, depending on the aspect ratio of the magnetic element [5-6]. Otherwise, the spin waves are manifested in the form of a set of radial and azimuthal modes spreading mainly outside the vortex core. The spin waves are described by integers (*n*, *m*), which indicate the number of the nodes of the dynamics magnetization component along the radial (*n*) and azimuthal (*m*) directions. Their eigenfrequencies, in the range of several gigahertz, are quantized due to strong pinning boundary conditions on the dot lateral edges [7-10]. Numerous studies have been carried out in these two types of dynamics, including the vortex gyrotropic motion excited by magnetic fields and spin-polarized currents [11-18], the stringlike gyrotropic waves [19], the azimuthal modes generated by a magnetic field parallel to the dot plane [20-21], and the radially symmetric modes generated by magnetic field pulse perpendicular to the dot plane [22-23].

Furthermore, the coupling between the vortex core and the spin-wave modes has inspired many interesting phenomenas. For example, coupling between the vortex core and the magnetostatic spin waves lifts the degeneracy of the clockwise and counterclockwise travelling magnons, leading to the zero-field splitting of the azimuthal spin-wave mode frequencies [24-27]. In turn, selective vortex core reversal can be achieved by taking advantage of the vortex polarity-dependent-frequency splitting of the azimuthal modes and even their interference [28-30]. In addition, resonance of radial spin waves can directly induce ultrafast vortex core reversal [31-33]. But to our knowledge, no investigations were reported regarding the gyrotropic motion directly excited by the spin waves. Here we present studies with the gyrotropic motion of magnetic vortex in nanodisks when excited by high frequency perpendicular magnetic fields. As we know, if the vortex is at rest in the disk center, perpendicular magnetic fields only excite the symmetric radial spin modes. These symmetric spin waves force the vortex core stay at the disk center, only lead to expansion and compression of the vortex core, no gyrotropic motion can appear. Therefore, our article describes the case of a shifted nonuniform magnetic vortex state, where an in-plane field is used to break the symmetry of the state, and an out-of-plane oscillating magnetic field is applied to excite the spin waves.

## Method and Results

Micromagnetic simulations were performed in a circular-shaped Permalloy nanodot (Fig. 1(a)), 200 nm in diameter, 5-nm thick by OOMMF code that employs the Laudau-Liftshitz-Gilbert equation [34]. The ground state for the disk is assumed to be a magnetic vortex with $(p, c)=(1, 1)$. In the simulations, the nanodisk was discretized into many small cells with each size of $2.5 \times 2.5 \times 10$ nm$^3$. The typical material parameters for Permalloy were used: the saturation magnetization $M_s = 8.6 \times 10^5$ A/m, the exchange constant $A = 1.3 \times 10^{-11}$ J/m, damping parameter $\alpha = 0.01$, and anisotropy constant $K_1 = 0$.

Many studies have concerned the spin waves excited in a shifted nonuniform magnetic vortex state [35-38]. However, the excitation fields are all in the disk plane, there is little understanding concerning the spin waves excited by perpendicular magnetic fields. Therefore, the first thing is examination the spin-wave modes in a shifted vortex excited by an out-of-plane magnetic field. We first apply an external in-plane field $H_y$ along the $y$ direction, the field is smaller than the field of magnetic saturation to make sure the stable state is still a vortex state. Under the effect of the bias field the vortex core is shifted away from the disk center and stabilized at an off-center position on the -$x$ axis after a period of time. Figure 1(b) shows the typical dependence of vortex core displacement from the disk center on $H_y$. Then, we apply a square wave pulse, 0.5ns in duration and 30 mT in strength, perpendicular to the disk plane to excite high frequency spin-wave modes. Fast Fourier transformation (FFT) of the $m_z$ component yields the spin-wave spectrum in a wide frequency range as a function of the bias field $H_y$, as shown in Fig. 1(c).

When the vortex core is at the center of the disk ($H_y=0$), the Fourier spectrum shows three spin-wave modes at 8.6, 12.6 and 16.6 GHz. The corresponding FFT power spatial distributions of $m_z$ are shown in the first row of Fig. 1(d). These patterns indicate that the three spin-wave modes corresponds to the radial modes with $n=1$, 2, and 3, respectively. Displacing the vortex core breaks the symmetry, the spectrum changes completely, some modes show very strong frequency shifts as a function of $H_y$. We observe different behaviors for the centered-vortex state (CV, $H_y<4$ mT), biased-vortex state (BV, $4<H_y<15$ mT) and edged-vortex state (EV, $H_y$>15 mT). In the CV state, three eigenmodes are observed with the eigenfrequencies almost independent of $H_y$. The corresponding FFT power spatial distributions of $m_z$ for the three excited modes at $H_y$ =1 mT are shown in Fig. 1(d). Little shift of the vortex core slightly breaks the cylindrical symmetry, but maintains the character of these modes like that at $H_y$= 0. All the three modes can be described as radial spin waves with the indices $n$=1, 2 and 3, respectively. When the bias field further increases, it makes the vortex strongly shift. In the BV state, the eigenmodes change obviously. The $n$=1 mode splits up into four distinct modes labeled p1, p2, p3, and p4. The $n$=2 mode (p5) does not splits, but its frequency decreases significantly when $H_y$ increases. The $n$=3 mode splits up into two mode labeled as p6 and p7. Frequency of the main branch (p7) decreases quickly. The mode p6 appears around $H_y$=8 mT, and gradually merges into p7. As examples, we show the FFT power spatial distributions of $m_z$ at the frequencies of these excited modes when $H_y$ =9 mT in Fig. 1(d). It is evident that the nodes symmetry along the radial direction is destroyed, but there are still some characteristics similar to the $H_y$=0 case. When the bias field is even further increased, the vortex core is close to the disk edge (EV state). In this state, besides the modes p1, p2, p3, p4, and p7, there are two additional modes labeled as EM-p9 and EM-p10 appear. As the FFT power spatial distributions of $m_z$ at $H_y$=20 mT (Fig. 1(d)) shown, these modes are completely different. The images show the formation of stripes in the region far from the vortex core, which is very similar to the case of uniform state [39]. Thus, it is not possible to associate these given modes with a corresponding one at zero field. Indeed, previous studies were mainly focused on the spin waves in the absence of biased magnetic field with the vortex core localized in the dot center. The spin wave modes ($n$, $m$) classification is usually applicable to the zero-field case [7-9]. For azimuthal modes, it was extended up to vortex nucleation field [37]. For our case, we think such a classification scheme is only applicable to the CV state.

Next, we show the relation between vortex gyrotropic motion and the spin-wave modes discussed above. We discuss two cases. The first one is the gyration of a shifted vortex induced directly by spin waves. In this case, we put the shifted vortex as the initial state, and apply a high frequency oscillating magnetic field $H_z = H_0 \sin(2\pi f_z t)$ perpendicular to the disk, while still retaining the value of $H_y$, where $H_0$ is the field amplitude, and $f_z$ is the field frequency, ranging from 1 to 20 GHz.

After applying $H_z$, the vortex is excited. We present the $y$ component of the magnetization $<m_y>$ average over the disk as a function of simulation time at different $f_z$ when $H_y$=10 mT in Fig. 2(a). Except the GHz-perturbation directly caused by the oscillating $H_z$, vortex gyrotropic motion can be infered by the MHz-oscilations of $<m_y>$. The lower frequency part of the corresponding Fourier spectrums shown in Fig. 2(b) just proves it. Moreover, we notice that $f_z$ has a big influence on the vortex gyrotropic motion. The amplitudes of $<m_y>$ are very small at 5.0 and 19.0 GHz. That is, the vortex core has no obvious motion. However, the amplitude increases with time when $f_z$ assumes 9.0, 11.2, and 15.6 GHz. This indicates that the gyration of vortex becomes obviously. Figure 2(c) shows the trajectory of vortex core for $f_z = 15.6$ GHz. The vortex core shows obvious spiral motion. Its distance from original position increases with time until reach a stable value (Fig. 2(d)). Here, we define the distance of vortex core from its initial position when $t$=20 ns as $d_v$ to measure the amplitude of gyrotropic motion. We also notice that the oscillations of $<m_y>$ are damaged when $f_z = 7.0$ and 8.0 GHz, which indicates that the magnetic state of the disk is no longer vortex. This distortion can be avoided when we reduce $H_0$ to 10 mT. Thus, in the following calculations, the value of $H_0$ are chosen to be 10 mT when $f_z < 10$ GHz in order to maintain the vortex state.

To clarify in more detail the relation between gyrotropic motion of the shifted vortex and the perpendicular oscillating field, we present the simulated $d_v$ in the plane of ($H_y$, $f_z$) in Fig. 2(e). There are a series of excite peaks appear, we also labeled them as p1, p2, .... It is very interesting to note, almost all the excited peaks can correspond the spin-wave modes labeled as the same symbols in Fig. 1(c). Just as the top panel in Fig. 1(c) and Fig. 2(e) shown, the marked peak frequencies in the $d_v$ curve can agree well with the eigenfrequencies in the spin-wave spectra, excepting the p1 mode. This agreement confirms that the gyrotropic motion of the shifted vortex is just excited by the spin waves, which induced by the perpendicular oscillating magnetic field. In addition, we find there is an excited gyration peak (p8) can not correspond to a certain spin-wave mode in the CV state. Its simulated FFT power image mode indicate that it should ascribed to the *n*=3 mode.

Then, we focus another condition, the influence of spin waves on the vortex gyration when vortex core is gyrating in the disk. Here, the bias field $H_y$ and the driving field $H_z$ are simultaneously applied to the disk, where the vortex core is initially in the disk center. Under the effect of $H_y$, the vortex core will gyrate around an off-centered equilibrium position. Meanwhile, the spin waves are excited by $H_z$. How will the spin waves impact on the gyration? Figure 3(a) shows the oscillations of magnetization <$m_y$> with simulation time at some specific frequencies when $H_y$=4 mT. For the no spin wave case ($H_z$=0), the amplitude of <$m_y$> clearly decreases with the simulation time. It is generally known that this is because of the existence of damping. The damping constant is chosen to be 0.01 in our simulation. However, if we include the effect of $H_z$, the oscillations of <$m_y$> have big changes. As examples, it shows anti-damping at 10.6, 12.2, and 16.6 GHz, but damping strengthen at 11.4 and 15.8 GHz.

In order to give a more extensive result, we calculate the effective damping constant $\alpha'$ for a series of $H_y$ and $f_z$, where the bias field $H_y$ varies from 0 to 13 mT. If $H_y$>13 mT, the vortex core may rotate out of the disk. As defined in Fig. 3(a), the effective damping constant $\alpha' = \alpha \left( \omega_I(H_y, f_z) / \omega_I(H_y, 0) \right)$, where $\omega_I(H_y, f_z) = (1/\Delta t) \ln(A(t_2)/A(t_1))$. The calculated $\alpha'$ in the plane of ($H_y$, $f_z$) are shown in Fig. 3(b). There are both anti-damping peaks (p2, p3, p4, p5, p7, p8) and damping strengthen peaks (a1, a2, a3, a4) appear. The anti-damping peaks are consistent with the spin-wave modes in the CV state, but they deviate obviously from the spin-wave modes at large $H_y$. The damping strengthen peaks can not correspond to any spin-wave modes in Fig. 1(c). This is because the spin-wave spectrum will be different when the vortex core gyrates in the disk. As described in Ref. 30, the eigenfrequencies strongly depend on the gyroexcitation amplitude.

## Discussions

To explain the above-described phenomenon, we start from the Thiele equation, which can explain the gyrotropic motion of vortex core well. For our system, it is given by [27, 40]

$$\mathbf{G} \times \frac{d\mathbf{X}}{dt} - D\frac{d\mathbf{X}}{dt} - \frac{dW}{d\mathbf{X}} + \dot{\mathbf{P}} = 0, \tag{1}$$

where $\mathbf{X}=(x,y)$ is the position vector of vortex core. The first term in Eq. (1) is the gyroforce. The gyrovector $\mathbf{G}=(M_s L/\gamma)\int \sin\theta(\nabla\theta \times \nabla\varphi)d^2\rho$, where $\gamma$ is the gyromagnetic ratio, $\theta$ and $\varphi$ are the polar and azimuth angles of magnetization. The second term is the damping force, where the damping tensor is expressed as $D=(\alpha M_s L/\gamma)\int (\nabla\theta)^2 + \sin^2\theta(\nabla\varphi)^2 d^2\rho$. The third term is the resorting force which is due to the change of the system energy. The total energy of the system can be written as $W_{tot}=W_m+W_{ex}+W_z$, where $W_m$, $W_{ex}$ and $W_z$ are the magnetostatic energy, the exchange energy, and the Zeeman energy. The fourth term is the force induced by the interaction of spin waves and vortex. We adopt the form given by Ref. 27.

Now, we analyze the effect of each force. Based on the simulation data, we calculate $\mathbf{G}$ and $D$ for different cases, including no spin wave case, anti-damping case, and damping strengthen case. After comparing these situations, we find the value of $|\mathbf{G}|$ is almost constant, having no relation with the spin waves. It agrees well with the results given by $\mathbf{G}=-G\hat{\mathbf{z}}$ with $G=2\pi p M_s L/\gamma$ [41]. The damping tensor $D$ have some changes. Compared to the no spin waves case, it increases and varies periodically with vortex position. However, the change of $D$ is small and it could not cause anti-damping gyrotropic motion of vortex core. For the force $\dot{\mathbf{P}}$, its influence on gyrotropic motion is just make perturbations because its frequency is much bigger than the gyrotropic frequency.

In order to study the effect of the third term, we consider the energies during the excitations. The effect of a fixed in-plane magnetic field has been widely studied. It makes the gyration center deviate from the disk center, but has no influence on the damping constant [42]. Moreover, a perpendicular magnetic field cannot affect the gyration damping of vortex core directly. Thus, we mainly focus on the exchange and demagnetization energy. Figure 4 shows the demagnetization energy and exchange energy of the whole disk with respect to the vortex core position during vortex gyrating at $H_y=$ 4 mT for different $H_z$. When $H_z=0$, there are no influence from spin waves, which corresponds to the well-known case, vortex gyration by an in-plane magnetic field [42]. For this condition, the exchange energy is almost independent of vortex core position (Fig. 4(b)), and the magnetostatic energy (Fig. 4(a)) is consistent with the formula $W_m=\kappa_0|\mathbf{X}|^2/2$, where $\kappa_0=10/9(M_s^2 L^2/R)$ is the stiffness coefficient [5]. Under the effect of $H_z$, spin waves are excited, the background magnetization is altered. We find the exchange energy (right panel in Fig. 4) become dependent on the vortex core position, whether anti-damping case (12.2 and 16.6 GHz) or damping strengthen case (15.8 GHz). Meanwhile, the relation between magnetostatic energy (left panel in Fig. 4) and **X** is no longer satisfy the above-formula.

The important character of the two energies is their distributions show frequency dependent. When $f_z$=15.8 GHz, both $W_m$ and $W_{ex}$ show the minimum when the vortex core located in the disk center, and they achieve the maximum when the vortex core is furthest from the disk center. This is similar to the case in Fig. 4(a), but its variation gradient gets bigger, so the damping is enhanced. In contrast, the energy distributions completely change for the anti-damping cases. Particularly, when $f_z$=16.6 GHz, the energies almost show the opposite distribution to the damping strengthen case. The different energy distributions of $W_m$ and $W_{ex}$ directly affect the term $dW/d\mathbf{X}$ in Eq. (1), and then influence the spiral trajectory of the vortex core.

In general, we confirm that the change of the energy distribution, including the exchange and the demagnetization energy, is mainly responsible for the gyration excitation and the damping change. The change of the energy distribution is induced by the spin waves, which is irrelevant to the vortex core polarity and the chirality. Indeed, the simulation results also confirm this idea. When we change the vortex polarity and chirality separately, we get the same results.

**Conclusion**

In summary, we found the vortex gyrotropic motion can be mediated by excitation of spin waves. An oscillating magnetic field was applied perpendicular to a nanodisk to excite spin waves. The excited spin waves alter the magnetization background of the disk. Then, the altered magnetization background changes the distribution of $W_m$ and $W_{ex}$ with respect to the vortex core position. This change on the one hand leads a shifted vortex to gyrate, and on the other hand changes the gyration damping when the vortex gyrating in the disk.

## Acknowledgments

This work was supported by the National Natural Science Foundation of China (Grant No. 11204026), and the Fundamental Research Funds for the Central Universities of Ministry of Education of China (Grant No. n130405011).

Fig. 1 (a) Schematic illustration of the nanodisk and its shifted vortex state, wherein vortex-core magnetization is upward and in-plane curling magnetization is counter-clockwise (white arrow). (b) Vortex core displacement from the disk center ($d_0$) as a function of $H_y$. (c) FFT power plots of the spin-wave spectra and its evolution as a function of $H_y$. The symbols p1, p2... label the spin-wave eigenmodes discussed in the text. Note that, for a better look, the values of FFT power for $f > 11$ GHz are magnified 15 times. A single spin-wave spectra like that for $H_y$=12 mT is shown on the top, where the excited peaks and their eigenfrequencies are marked. (d) FFT power spatial distributions of $m_z$ for the spin-wave modes at $H_y$=0, 1 (CV), 9 (BV), and 20 mT (EV).

Fig. 2 Spin waves-induced gyration of a shifted vortex. (a) Variation of <$m_y$> of the shifted vortex ($H_y$=10 mT) as a function of simulation time when excited by a perpendicular oscillating magnetic field with the amplitude $H_0$=30 mT, where $f_z$=5.0, 7.0, 8.0, 9.0, 11.2, 15.6 and 19.0 GHz, respectively. (b) The corresponding FFT power spectrum of <$m_y$> given in (a). (c) The trajectory and (d) the distance from its original position ($d$) of vortex core variation with simulation time of vortex core for $f_z = 15.6$ GHz. (e) Color plot of $d_v$ with respect to $H_y$ and $f_z$. A single $d_v$ curve like that for $H_y$=12 mT is shown on the top, where the excited peaks and their frequencies are marked.

Fig. 3 Vortex gyration mediated by spin waves. (a) Variation of <$m_y$> as a function of simulation time excited by a bias field $H_y$=4 mT when $H_z$=0, and $H_z$=30 mT with $f_z$=10.6, 11.4, 12.2, 15.8, and 16.6 GHz, respectively. (b) Color plot of the effective damping constant $\alpha'$ with respect to $H_y$ and $f_z$. The symbols p1, p2... label the anti-damping modes, and the symbols a1, a2... label the damping strengthen modes.

Fig. 4 Demagnetization energy ($W_m$) and exchange energy ($W_{ex}$) during vortex gyrating at $H_y$=4 mT when $H_z$=0, and $H_z$=30 mT with $f_z$=12.2, 15.8, and 16.6 GHz, respectively. The red balls show the x-y projection of the vortex core positions.

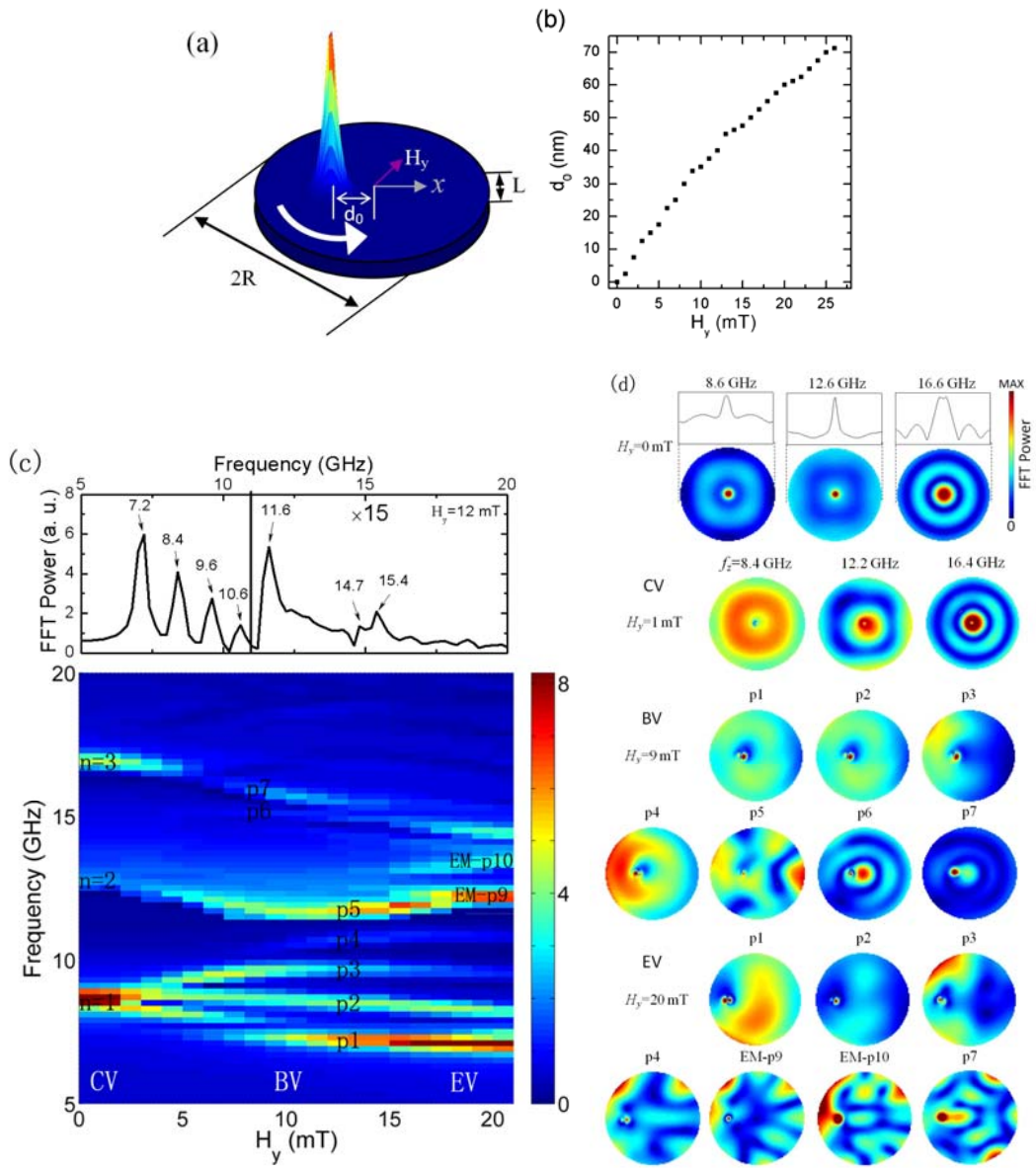

Figure 1

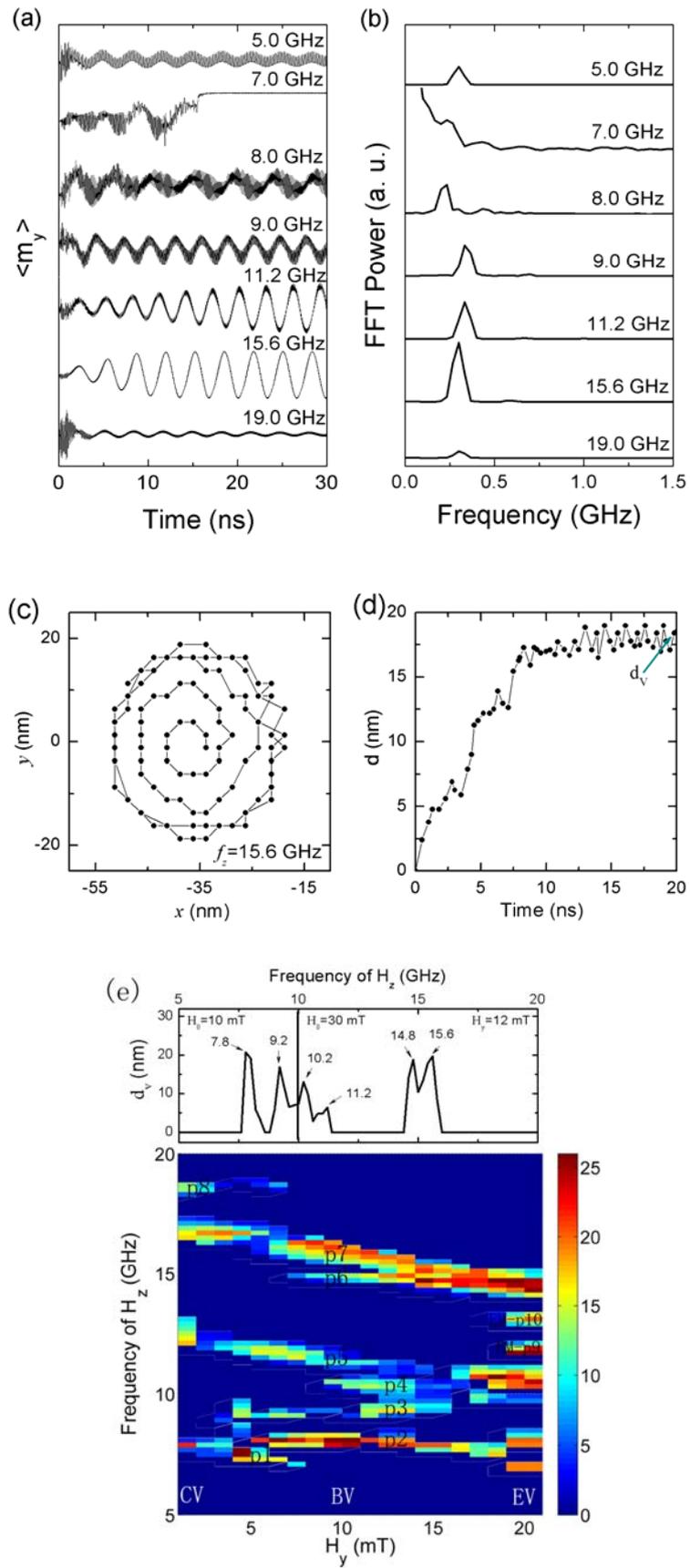

Figure 2

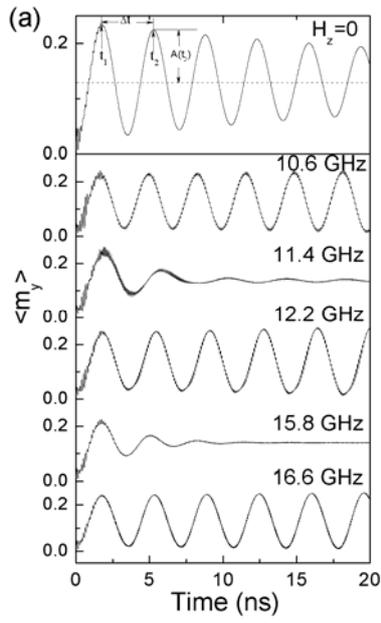
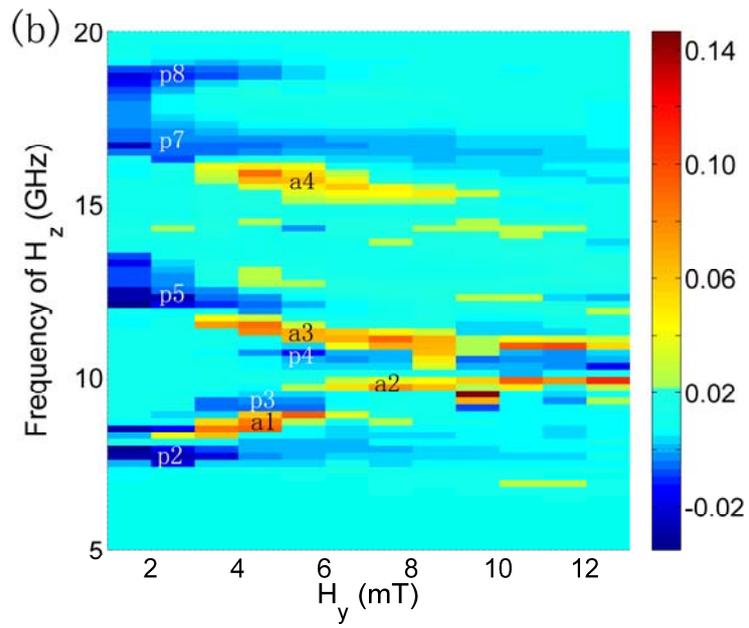

Figure 3

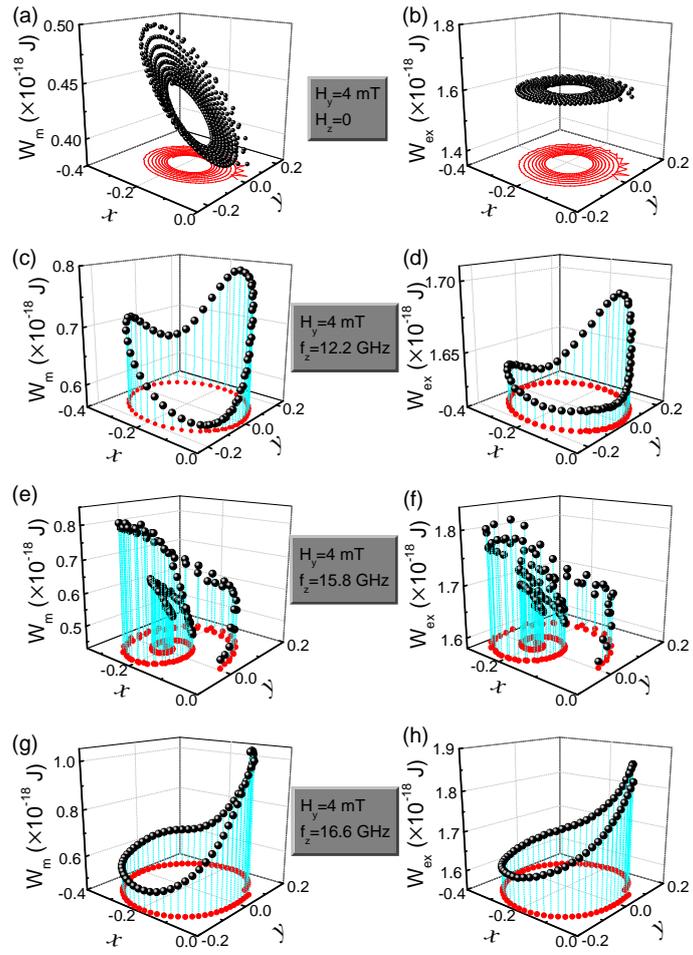

Figure 4